\documentclass{article}[16pt]
\usepackage[utf8]{inputenc}
\usepackage{graphicx}
\usepackage{lscape}
\begin{document}
\begin{center}

{\bf \Large SURFACE PHOTOMETRY OF DWARF IRREGULAR GALAXIES IN DIFFERENT ENVIRONMENTS}

\bigskip

{\bf Sharina~M.E.}

\bigskip

{\bf Special Astrophysical Obseratory of the Russian Academy of Sciences, Nizhnyi Arkhiz, Russia\\ e-mail: sme@sao.ru}
  \end{center}
  
\bigskip

{\em Abstract 
Surface photometry data on 90 dwarf irregular galaxies (dIrrs) in a wide vicinity of the Virgo cluster and 30 isolated dIrrs are presented. Images from the Sloan Digital Sky Survey (SDSS) are used. The following mean photometric characteristics (color and central surface brightness) are obtained for objects in the two samples: $(V-I)_0=0.75~mag~(\sigma=0.19~mag)$, $(B-V)_0=0.51~mag~(\sigma=0.13~mag)$, $SB_V=22.16~mag/sq.arcsec~(\sigma=1.02~mag/sq.arcsec)$  for the dIrrs in the vicinity of the Virgo cluster and $(V-I)_0=0.66~mag~(\sigma=0.43~mag)$, $(B-V)_0=0.57~mag~(\sigma=0.16~mag)$, $SB_V=22.82~mag/sq.arcsec~(\sigma=0.73~mag/sq.arcsec)$  for the isolated galaxies. The mean central surface brightness for the isolated galaxies in this sample is lower than the mean central surface brightness for the dIrrs in a denser environment. The average color characteristics of the dIrrs in the different environments are the same to within $\sim$0.2 mag.
} \\

\bigskip

{\bf 1. Introduction}

\bigskip
In this paper the photometric and structural parameters of a large sample of irregular dwarf galaxies (dIrrs) are
obtained in different environments within roughly 20 Mpc. Here, by dIrr we mean a morphological type of dwarf galaxies, intermediate between dwarf elliptical or dwarf shperoidal galaxies, on the one hand, and blue compact dwarf galaxies, on the other. The former have lost gas, are elliptical or spheroidal in shape and have no recent star formation. The latter are fully involved in star formation.
 The radial velocities of the galaxies relative to the centroid of the Local group are $V_{LG}<3500$~km/s.
 The velocities are taken from the HyperLeda\footnote{http://leda.univ-lyon1.fr/} data base [1].
 The exact photometric distances to the objects in the sample are not known, except for 4 of the galaxies: UGC7150,
VCC530, UGC7784, and VCC2037 [2].
Our sample of dIrrs in the vicinity of the Virgo cluster was constructed
in fairly random fashion. It is part of a huge list of such objects and includes galaxies in projection on the virialized
zone of the cluster and in the $\sim$12-degree infall zone of the cluster, as well as more distant objects in projection on
the sky. The isolated dIrrs in our sample are taken from Ref. 3.
 These are galaxies which can be seen in CCD images in the SDSS with radial velocities $V_{LG} < 3500$~km/s 
 and relative differences $>500$~km/s 
  in their velocities with nearest neighbors projected on the sky (projected distances  $>500$~kpc). 
  The morphological types of the dwarf galaxies are
given in Ref. 3, along with a list of major work on the detection of dwarf galaxies with low surface brightness in
the Local volume. The objects for which photometric data are given in this paper and for which the designations
begin with KK, KKR, and KKH were first discovered by Karachentseva, Karachentsev, et al. [4-6] in charts from the
Palomar Atlas of the Sky. The KDG galaxies were discovered by Karachentseva [7]. Other abbreviations encountered
among the objects of our study and found in the literature include the following: UGC (Uppsala General Catalogue)
[8], PGC (Catalogue of Principal Galaxies, Lyon-Meudon Extragalactic Database) [9,10,1], MGC (Millennium Galaxy
Catalogue) [11], VCC (Virgo Cluster Catalogue) [12], the VLSB catalogs of galaxies with low surface brightness of
Schombert, et al.,from 1988-1997 [13-15], and the AGCArecibo Legacy Fast ALFA Survey [16].

The main properties of dwarf galaxies of different morphological types within 10 Mpc are summarized in Refs.
17 and 18. Do the photometric properties of dIrrs depend on their environment? We attempt to answer this question
here on the basis of data obtained for objects whose spatial distribution extends beyond the confines of the Local
volume.

\bigskip                                                                                                                    
                                                                                                                            
{\bf 2. Photometry}                                
                                                                                                                            
\bigskip

The surface photometry process was analogous to that in Refs. 18 and 19. The SDSS provides images in the
u, g, r, i, and z bands that have been through a preliminary reduction process (dark current and system electronic zero
subtraction, splitting into flat field). We used images in the g, r, and i bands for the photometry. Photometric
coefficients are given in the SDSS site for each image for conversion of the photometric results from the instrumental
system into the standard SDSS system. The {\it SURFPHOT} program package was used to carry out the photometry. This
is part of MIDAS (Munich Image Data Analysis System) [20] a large package of programs for analysis of astronomical
data developed at ESO. MIDAS programs were used to mask background stars and galaxies. The
 {\it FIT/BACKGROUND} program was used to rectify and subtract the sky background from the initial images. The program 
 {\it FIT/ELL3} for
inscribing ellipses was used to search for the centers of the galaxies and model the intensity distribution over the
area of an object. The flux was integrated in the resulting elliptical apertures and the azimuthally averaged surface
brightness calculated. Additional photometry was carried out using programs from the MIDAS package in concentric
rings of thickness 1 pixel with a center determined in advance during inscription of the ellipses. The photometric
results were converted from the SDSS photometric system to the standard Johnson-Cousins photometric system using
the empirical formulas from Ref. 21.

During photometric processing it was found that most of the objects being examined have the following
structure: a stellar disk that dominates the intensity with a brightness that falls off smoothly from the center to the
edge, with star-formation complexes randomly distributed over the object superimposed on it. In these cases the disk
component is well described by a sequence of model ellipses. After this model is subtracted, the weak star-formation
complexes remain. The individual dIrrs have an extremely weak gradient of the surface brightness. If these galaxies
are positioned en face or if the angle between the polar axis of an object and the line of sight $incl > 45^o$, then the
growth curves and profiles of the surface brightness are best constructed using concentric rings. Objects of this kind
with  $incl < 45^o$ were not examined in the present work. Photometry was carried out for all the dIrrs in the sample
with both elliptical and circular apertures. The resulting brightness profiles and growth curves were compared. It
should be noted that for all the objects in the sample the photometric parameters determined by these two methods
are essentially the same.

Figure 1 illustrates the process of inscribing ellipses for the galaxy KDG104. Figure 2 shows the photometric
results for this galaxy converted to the standard Johnson-Cousins photometric system: growth curves in the B, V, R,
and I bands, profiles of the surface brightness in these bands, and the radial dependences of the integral and
differential color characteristics $B-V$, $R-I$. 
 Similar figures for the other galaxies in the sample are shown at the
ftp site of the Special Astrophysical Observatory of the Russian Academy of Sciences\footnote{ftp://ftp.sao.ru/pub/sme/dIrrs120/}. 
Tables with profiles of the surface brightness in all the bands are given there for each of the objects that
were studied.

The errors in surface photometry should be noted. Random errors are determined by the deviation of the
intensity counts within the aperture from the mean relative to the corresponding value for the sky background
surrounding the object. 
The range of the oscillations in surface brightness ($\rm SB$)
relative to the overall dependence of $\rm SB$ on R (the radius of the galaxy) (Fig. 2, right hand frames) corresponds in
magnitude to the random photometric errors at a given distance from the center of the galaxy. The accuracy with
which the overall background level around the galaxy is specified is also determined by the random fluctuations in
background intensity in the region being studied after masking of background objects. The systematic photometric
errors include the accuracy of the zero points for conversion from the instrumental into the standard system and
internal, unaccounted-for absorption in the galaxies being studied. The combined (random+systematic) errors in the
integral stellar magnitudes depend on the brightness of the object being examined and equal $0.2-0.3$  for the
objects in the sample. The errors in measuring the surface brightness are given for each galaxy and for each point
in the surface brightness profile (tables at the ftp site).
\begin{figure*}
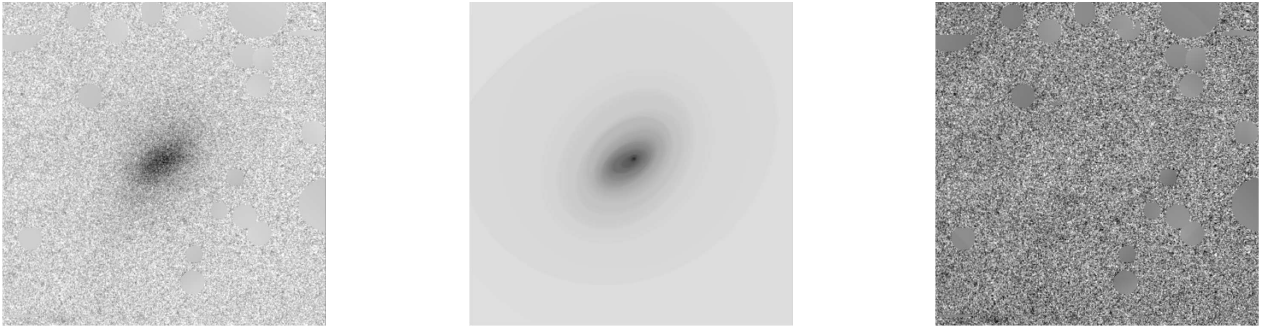

\begin{minipage}[h]{0.27\linewidth}
\includegraphics[scale=0.27]{pictKDG104g.ps}
\end{minipage}
\hfill
\begin{minipage}[h]{0.27\linewidth}
\includegraphics[scale=0.27]{ellKDG104g.ps}
\end{minipage}
\hfill
\begin{minipage}[h]{0.27\linewidth}
\includegraphics[scale=0.27]{resKDG104g.ps}
\end{minipage}
\caption{Illustrating the process of inscribing ellipses in an SDSS image of the galaxy KDG104
(g band) (from left to right): (1) original frame with background objects removed and the sky
background subtracted; (2) the model; (3) the object minus the model. The frame size is $2'\times2'$.}
 \end{figure*}
 
\begin{figure*}
\begin{minipage}[h]{0.47\linewidth}
\includegraphics[scale=0.27,angle=-90]{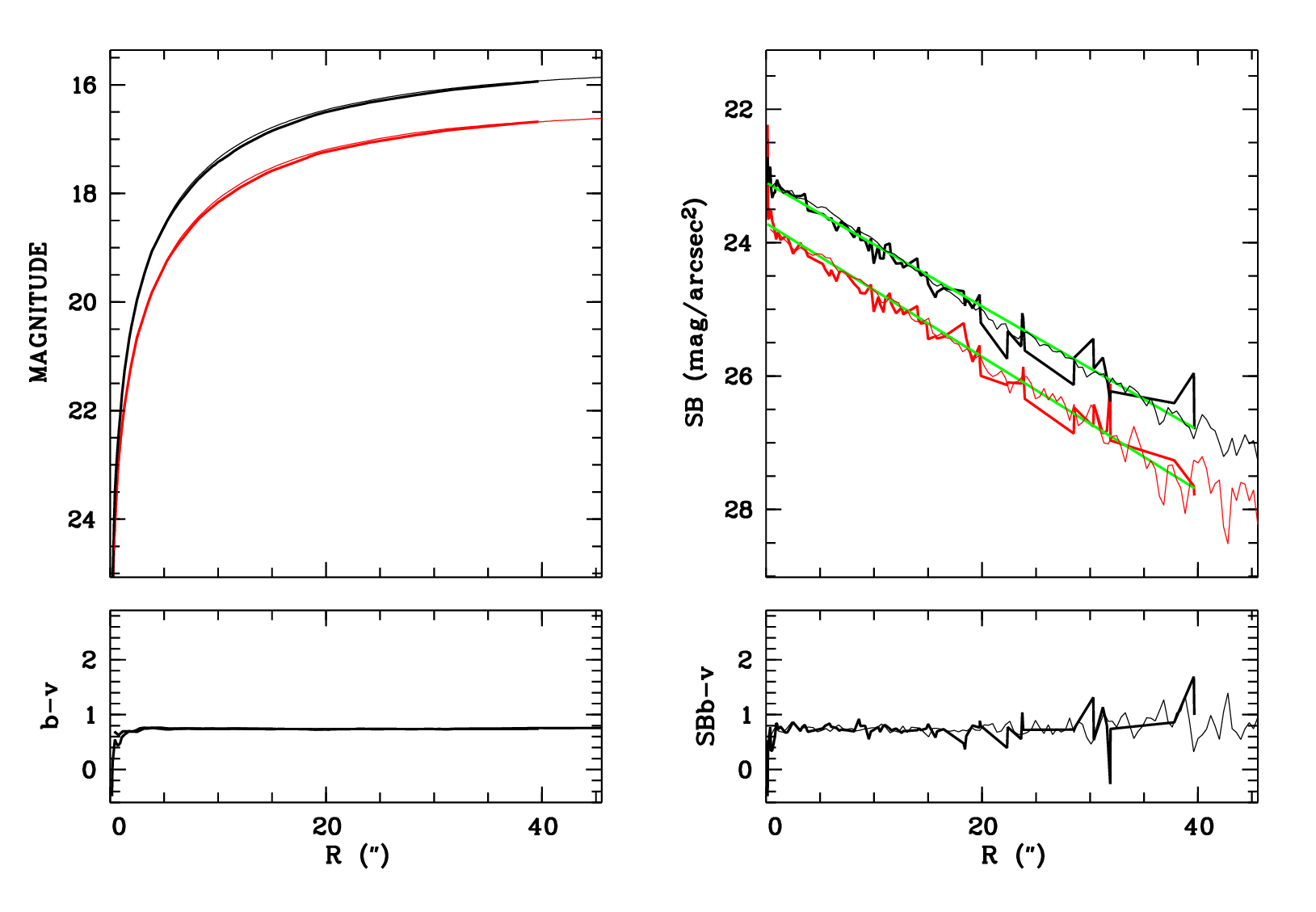}
\end{minipage}
\hfill
\begin{minipage}[h]{0.47\linewidth}
\includegraphics[scale=0.27,angle=-90]{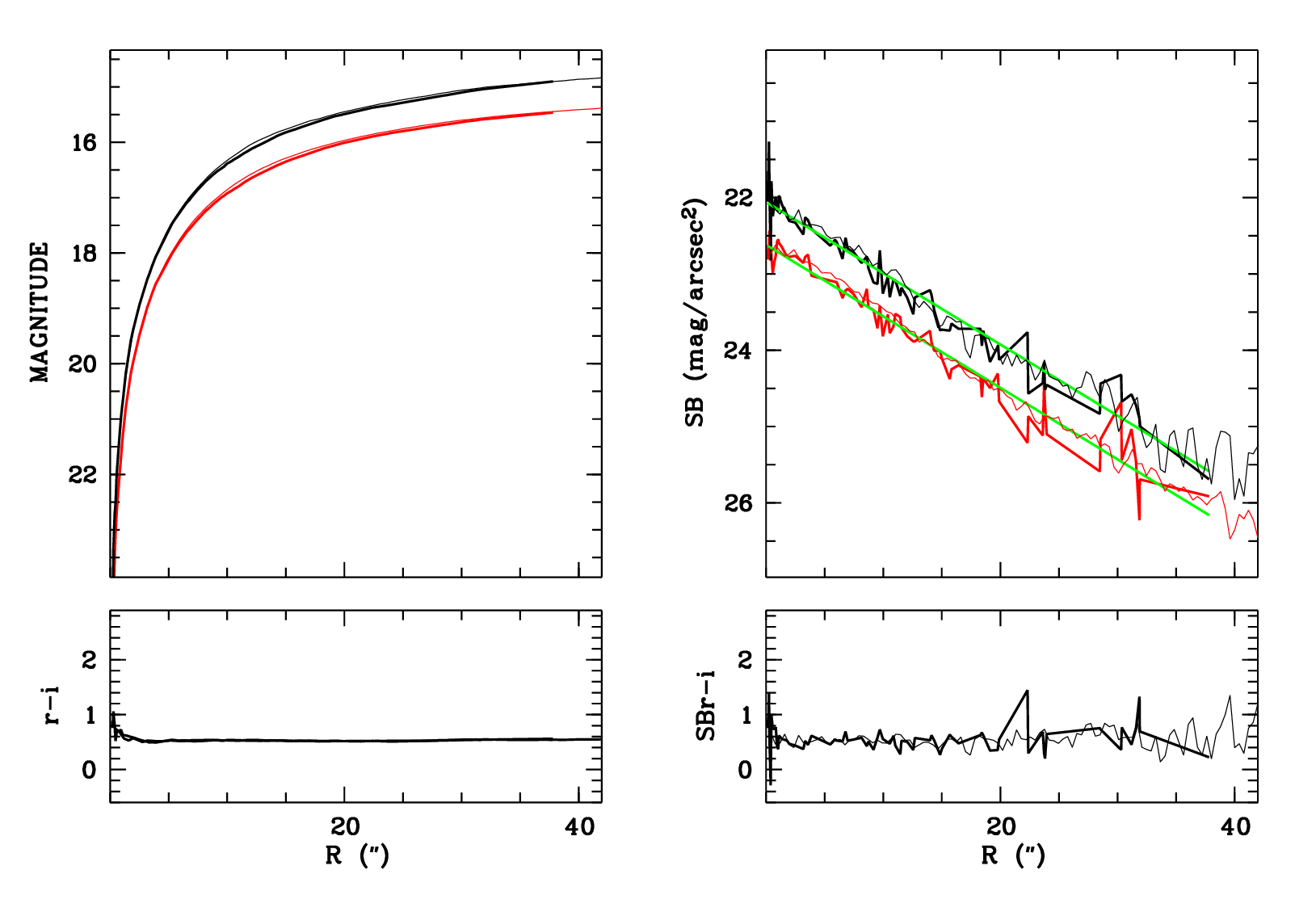} 
\end{minipage}
\caption{Results of surface photometry of the galaxy KDG104 uncorrected
for absorption of light in the galaxy, in the B and V (top) and R and I
(bottom) bands: the growth curves and corresponding radial dependences
(left frames of each of the figures) of the colors $\rm B-V$ and $\rm R-I$; profiles of
the surface brightness in the same bands and the corresponding radial
dependences of the colors (right frames). The thin curves show the
photometry with circular apertures and the thick curves, with elliptical
model apertures. The straight lines are approximations of the surface
brightness profiles by exponential functions.}
\end{figure*}

\bigskip                                                                                                                    
                                                                                                                            
{\bf 2. Results}                                
                                                                                                                            
\bigskip
Tables 1 and 2  show the results
of the surface photometry in the B, V, R, and I bands of the standard broadband Johnson-Cousins photometric system
and the model parameters for the surface brightness profiles of the galaxies obtained using the exponential function [22]: 
$SB(r)=SB_0+1.086(r/h)) $, where $SB_0$  is the central surface brightness and $h$ is the exponential length scale.
The columns of Tables 1 and 2 contain the following data: (1) the designation of the galaxy, (2) the right ascension
(format: $hh$ $mm$ $ss$) and declination (format: $^o$, $'$, $''$) at epoch J2000.0; (3) the galactic absorption in the V band
(in stellar magnitudes) (row 1), the ratio of the lengths of the minor and major axes of the galaxy (row 2), and the
velocity relative to the centroid of the Local group (in km/s) (row 3); (4) the integrated B, V, R, and I stellar magnitudes
(in stellar magnitudes); (5) the integrated colors $(B-V)_0$, $(R-I)_0$ (in stellar magnitudes), and corrected for galactic
absorption; (6) the observed central surface brightness (in $\rm mag/sq.arcsec$); (7) integrated stellar magnitudes corresponding
to the level of surface brightness of the azimuthally averaged profile, 25 mag/sq.s; (8) average surface brightnesses
within the level of surface brightness of the azimuthally averaged profile 25 $\rm mag/sq.arcsec$; (9) the effective radii (in
ang.s) encompassing half the luminosity of the galaxy, measured from the luminosity growth curve of the galaxy;
(10) the exponential scaling lengths (in ang.s); (11) the integrated stellar magnitudes corresponding to the effective radii
in each band (mag); (12) median surface brightnesses within the effective radii ($\rm mag/sq.arcsec$); (13) the model central
surface brightnesses ($\rm mag/sq.arcsec$); (14) the minimum (row 1) and maximum (row 2) radii ($''$),  within which the
azimuthally averaged surface brightness was modelled by an exponential law.

Our measured full diameters of the dIrrs in the sample (isolated+Virgo) are compared in Fig. 3 with the
diameters for these objects at a surface brightness level of 25 $\rm mag/sq.arcsec$ in the $B$-band taken from the HyperLeda data base. 
 It can be seen that our diameters are an average of two times greater than those from HyperLeda at an isophot level of
 25 $\rm mag/sq.arcsec$. This result can be expected given the depth of our surface photometry.

The averaged photometric characteristics corrected for galactic extinction have been compared [23] for dwarf
galaxies of various morphological types in the Local group ($Dist<1$~Mpc) and the Local volume $Dist<10$~Mpc 
beyond the confines of the Local group: absolute magnitudes, colors, and average central surface brightness in the
V band corrected for absorption of light in the galaxy ($SB_{V}$).
It turns out that the average photometric properties of
the dIrrs in the Local universe are very similar: $M_V=-12.93~mag~(\sigma=0.2~mag)$, 
$(V-I)_0=0.73~mag~(\sigma=0.2~mag)$, $(B-V)_0=0.47~mag~(\sigma=0.2~mag)$, $\rm SB_{V}=22.4 mag/sq.arcsec~(\sigma=0.7)$. 
 This means that the average sizes, masses, and densities of the stars
and star-formation complexes, and the levels of star formation in galaxies of this morphological type are roughly the
same in the Local group and in the Local volume containing groups of galaxies and voids with isolated objects in
them.

We now compare the photometric properties of our two samples of galaxies with the properties of the galaxies
in the Local volume. The average colors and central surface brightnesses of the galaxies in the vicinity of the Virgo
cluster are $(V-I)_0=0.75~mag~(\sigma=0.19~mag)$, $(B-V)_0=0.51~mag~(\sigma=0.13~mag)$, $SB_V=22.16~mag/sq.arcsec~ (\sigma=1.02 mag/sq.arcsec)$. 
The isolated dIrrs in the sample
have the following average characteristics: $(V-I)_0=0.66~mag~(\sigma=0.43~mag)$, $(B-V)_0=0.57~mag~(\sigma=0.16~mag)$,
 $SB_V=22.82~mag/sq.arcsec~(\sigma=0.73~mag/sq.arcsec)$.
It can be seen that the average colors corrected for absorption are similar for dIrrs in the Local universe
and in its surroundings. That is, the average star formation rate for the dIrrs in different environments is roughly the
same. The average central surface brightnesses for the isolated objects in the sample are significantly lower (by
roughly a factor of 2) than for the dIrrs in the Local group, in the Local volume beyond the confines of the Local
group, and for the dIrrs in our sample in the vicinity of the Virgo cluster. This means that, on the average, the isolated
dIrrs in the sample have lower gradients of the brightness from the center to the edge. If the average masses of the
dIrrs in different environments are approximately the same, then the lower brightnesses of the galaxies in the center
which we have obtained may mean that the sizes of the isolated dIrrs in the sample are larger by a factor of 2 on
the average than for objects of this morphological type in a denser environment.
\begin{figure*}
\begin{center}
\includegraphics[width=0.56\textwidth,scale=2]{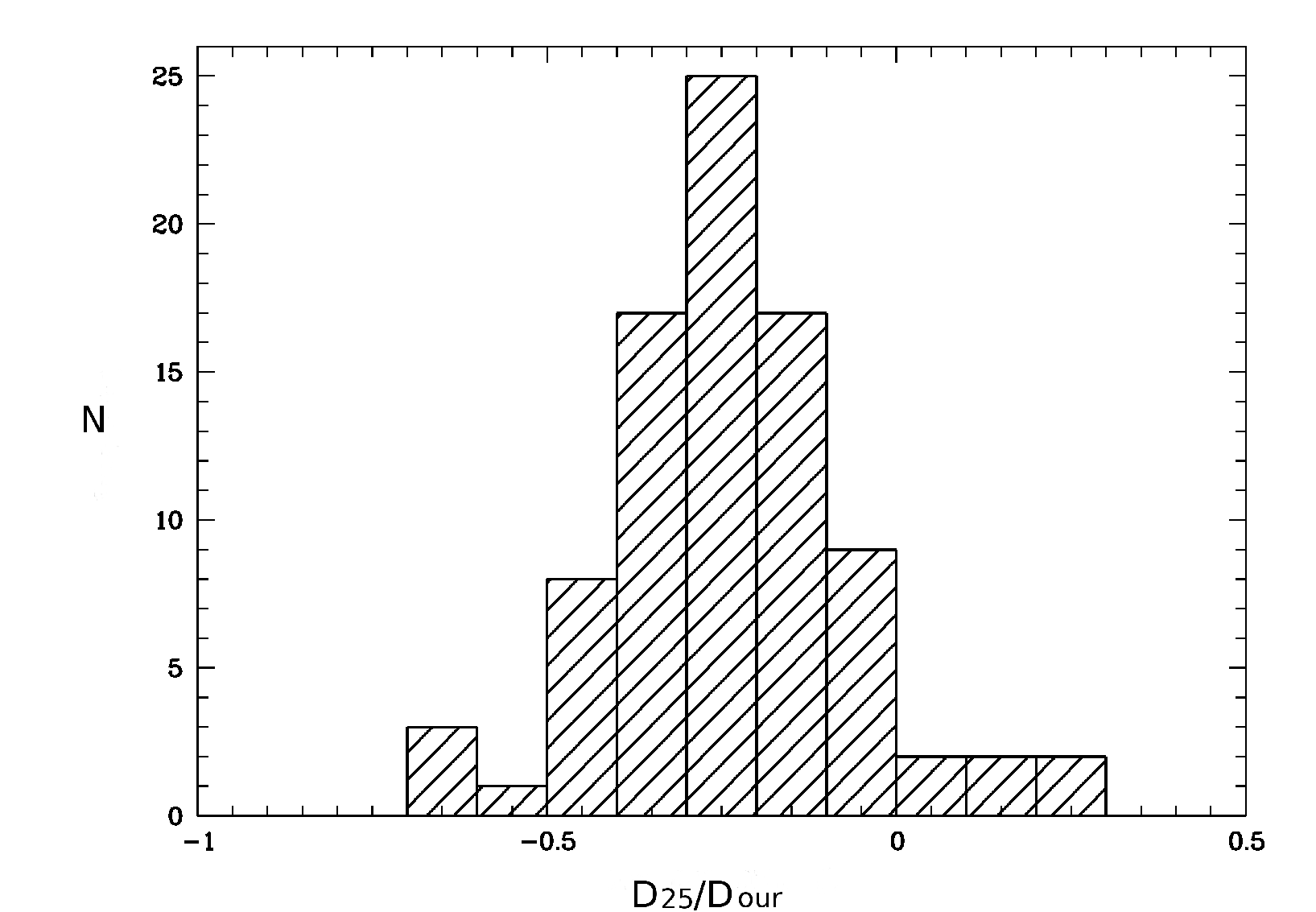} 
\caption{The distribution of the ratio of the diameters of the dIrrs from HyperLeda at a level of $SB=25$~$\rm mag/sq.arcsec$.
 to our measured total diameters of these galaxies.}
 \end{center}
 \end{figure*}

\bigskip

{\bf 5. Conclusion}

\bigskip
Results on surface photometry and measurements of the structural parameters have been presented here for
30 isolated dIrrs from Ref. 3 and 90 dIrrs in the vicinity of the Virgo cluster. The surface brightness profiles of the
galaxies are approximated by an exponential law.

We have found that the average color indices for the isolated objects in the sample and galaxies in the vicinity
of the Virgo cluster are in good agreement with those for dIrrs in the Local group and Local volume. The central
surface brightnesses of the isolated galaxies are significantly lower than those of the dIrrs in a denser environment.

As noted in the Introduction, the exact photometric distances to most of the galaxies in the sample are not
known. We found that the isolated dIrrs in our sample are larger on the average than objects of this morphological
type in a denser environment. Measurement of the distances to the objects in the sample will help detect new
representatives of the family of so-called ultra-diffuse galaxies [24] characterized by effective radii
$R_{eff} = 1.5 \div4.6$~kpc, absolute stellar magnitudes typical of dwarf galaxies, and central surface brightnesses fainter than
24~$\rm mag/sq.arcsec$ in the visible.

\bigskip
  {\bf Acknowledgements}
  This study was supported by a grant from the Russian Science Foundation (project No. 14-12-00965). The
author thanks D. I. Makarov and I. D. Karachentsev for providing the list of galaxies in the vicinity of the Virgo cluster.
Data from the Sloan Digital Sky Survey (SDSS) have been used in this work. The SDSS is a joint project of the
University of Chicago, Fermilab, the Institute for Advanced Study, the Japanese participant group, Johns Hopkins
University, the Max Planck Institute of Astronomy (MPIA), the Max Planck Institute for Astrophysics (MPA), New
Mexico State University, Princeton University, the US Naval Observatory, and the University of Washington. The
Apache Point Observatory, where the SDSS telescopes are located, is operated by the Astrophysical Research Con-
sortium (ARC). This work was done using data from the HyperLeda data base (http://leda.univ-lyon1.fr).

\bigskip

  {\bf Bibliography}
  
\bigskip
1.   Makarov D.I., Prugniel P., Terekhova N., Courtois H., Vauglin I., Astron. Astrophys. 570, 13 (2014).

2.   Karachentsev I. D., Tully R.B., Wu P.-F., Shaya E.J., Dolphin A.E., Astrophys. J., 782, 4 (2014).
  
3.   Karachentseva V. E., Karachentsev I. D., Sharina M. E., Astrophysics, 53, 462 (2010).
  
4.   Karachentseva V. E., Karachentsev I. D., Astron. Astrophys. Suppl. Ser., 127, 409 (1998).
  
5.   Karachentseva V. E., Karachentsev I. D., Richter G. M., Astron. Astrophys. Suppl. Ser., 135, 221 (1999).
  
6.   Karachentsev I. D., Karachentseva V. E., Huchtmeier W. K., Astron. Astrophys., 366, 428 (2001).
  
7.   Karachentseva V.E., Soobshch. Byurakan Obs., 39, 6 (1968).
  
8.   Nilson P., Acta Univ. Uppsala, ser. V, vol.1 (1973).
  
9.   Paturel G., Fouque P., Bottinelli L., Gouguenheim L. Astron. Astrophys. Suppl. Ser., 80, 299 (1989).
  
10.  Paturel G., Petit C., Prugniel P., Theureau G., Rousseau J., Brouty M., Dubois P., Cambresy L., Astron. Astrophys., 412, 45 (2003).
  
11.  Liske J., Lemon D. J., Driver S. P., Cross N. J. G., Couch, W. J., Mon. Not. Roy. Astron. Soc., 344, 307 (2003).

12.   Bingelli B., Sandage A., Tammann G.A., Astron. J., 90, 1681 (1985).
  
13.   Schombert J.M., Bothun G.D., Astron. J. 95, 1389 (1988).
  
14.   Schombert J.M., Bothun G.D., Schneider S.E., McGaugh S.S., Astron. J., 103, 1107 (1992).
  
15.   Schombert J.M., Pildis R.A., Eder J.A. Astrophys. J. Suppl. Ser., 111, 233 (1997). 

16.   Haynes M. P., Giovanelli R., Martin A. M., et al., Astron. J., 142, 170 (2011).
  
17.   Karachentsev, I. D., Makarov D.I., Kaisina E.I., Astron. J., 145, 101 (2013).
  
18.   Sharina M.E., et al., Mon. Not. R. Astron. Soc. 384, 1544 (2008).
    
19. Sharina M.E., Il’ina E.A., Astron. Nachr. 334, 773 (2013).

20. K. Banse, P. Crane, P. Grosbol, et al., The Messenger 31, 26 (1983).

21. Jordi K., Grebel E. K., Ammon K., Astron. Astrophys., 460, 339 (2006).

22. de Vaucouleurs G., in Flugge S., ed., Handbuch der Physik 53. Springer-Verlag, Berlin, p. 275 (1959).

23.   Sharina M.E., Karachentseva V.E., Makarov D.I. in de Grijs, R. ed. IAU Symposium 289, Advancing the Physics of Cosmic Distances, 
    Cambridge Univ. Press, Cambridge, p. 236 (2013).

24. van Dokkum P. G., Abraham R., Merritt A., et al. 2015, Astrophys. J. Letters, 798, L45 (2015).

\begin{table}
\begin{center}
\vspace{-1.0cm}
\scriptsize
\caption{\label{tab:properties} Photometric Properties of the Isolated dIrrs in the Sample.}
\begin{tabular}{l|c|c|c|c|c|c|c|c|c|c|c|c|r}
\hline\hline 
Name   & RA(2000)   &  $A_V$ & $B$&$(B-V)_0$&$SB_{B_0}$&$B_{25}$&$SB_{B_{25}}$ &  $Re_B$&  $h_B$  &$B_e$& $SB_{B_e}$&  $SB_{B_c}$&$R_{min}$       \\
       & Dec.(2000) & $b/a$&   $V$&$(R-I)_0$&$SB_{V_0}$&$V_{25}$&$SB_{V_{25}}$ &  $Re_V$&  $h_V$  &$V_e$& $SB_{V_e}$&  $SB_{V_c}$&$R_{max}$       \\ 
       &            & $V_{LG}$&$R$ &        &$SB_{R_0}$&$R_{25}$&$SB_{R_{25}}$ &  $Re_R$&  $h_R$  &$R_e$& $SB_{R_e}$&  $SB_{R_c}$&      \\
       &            &         & $I$ &       &$SB_{I_0}$&$I_{25}$&$SB_{I_{25}}$ &  $Re_I$&  $h_I$  &$I_e$& $SB_{I_e}$&  $SB_{I_c}$&          \\ 
  {\tiny (1)}  &    {\tiny (2)}     &   {\tiny (3)}&  {\tiny (4)}  &  {\tiny (5)}   & {\tiny (6)}  &  {\tiny (7)}      & {\tiny (8)}       &  {\tiny (9)}  &  {\tiny (10)} &  {\tiny (11)}   &   {\tiny (12)}       &  {\tiny (13)}  & {\tiny (14)} \\ \hline
KKH14  & 02 45 02.9 & 0.63 & 16.31& 0.23  &24.52& 18.18& 24.99 &26.74 & 16.06 &17.08 &  25.40&  24.30&  0.79     \\
       & 32 09 42.0 & 0.96 & 15.89& 0.61  &24.01& 17.00& 24.63 &24.24 & 14.96 &16.58 &  24.74&  23.66&  53.46    \\
       &            & 1748 & 15.58&       &23.67& 16.21& 24.22 &23.89 & 15.49 &16.14 &  24.23&  23.37&          \\
\hline\hline   
\end{tabular}                                                                                                                                              
\end{center}                                                                                                                
\end{table} 
 \setcounter{table}{0}
\begin{table}
\begin{center}
\vspace{-1.0cm}
\scriptsize
\caption{ Photometric Properties of the Isolated dIrrs in the Sample (continued).}
\begin{tabular}{l|c|c|c|c|c|c|c|c|c|c|c|c|r}
\hline\hline 
Name     & RA(2000)    &  $A_V$&   $B$ &$(B-V)_0$  &$SB_{B_0}$&$B_{25}$&$SB_{B_{25}}$ &$Re_B$&  $h_B$  & $B_e$ & $SB_{B_e}$& $SB_{B_c}$& $R_{min}$       \\
         & Dec.(2000)  & $b/a$&   $V$  &$(R-I)_0$  &$SB_{V_0}$&$V_{25}$&$SB_{V_{25}}$ &$Re_V$&  $h_V$  & $V_e$ & $SB_{V_e}$& $SB_{V_c}$& $R_{max}$       \\ 
         &             &$V_{LG}$& $R$  &           &$SB_{R_0}$&$R_{25}$&$SB_{R_{25}}$ &$Re_R$&  $h_R$  & $R_e$ & $SB_{R_e}$& $SB_{R_c}$&           \\
         &             &      &   $I$  &           &$SB_{I_0}$&$I_{25}$&$SB_{I_{25}}$ &$Re_I$&  $h_I$  & $I_e$ & $SB_{I_e}$& $SB_{I_c}$&           \\  
    {\tiny (1)}  &    {\tiny (2)}     &   {\tiny (3)}&  {\tiny (4)}  &  {\tiny (5)}   & {\tiny (6)}  &  {\tiny (7)}      & {\tiny (8)}       &  {\tiny (9)}  &  {\tiny (10)} &  {\tiny (11)}   &   {\tiny (12)}       &  {\tiny (13)}  & {\tiny (14)} \\ \hline
VLSB     & 13 55 58.0 & 0.08 & 17.98& 0.75  &24.28&19.39& 24.63  &10.63 &  6.77 &18.62 &  24.94&  24.06&  0.40      \\ 
721-10   & 08 59 29.0 & 0.64 & 17.20& 0.54  &23.82&18.20& 24.24  &11.89 &  7.32 &17.92 &  24.44&  23.53&  20.99     \\ 
         &            & 1165 & 16.68&       &23.31&17.50& 23.98  &12.95 & 7.89  &17.44 &  24.11&  23.18&           \\
\hline\hline                                                                                                                          
\end{tabular}                                                                                                                                              
\end{center}                                                                                                                
\end{table}                                                                                                                 
\begin{table}
\begin{center}
\vspace{-1.0cm}
\scriptsize
\caption{Photometric Properties and Velocities of the dIrrs in the Vicinity of the Virgo
Cluster}
\begin{tabular}{l|c|c|c|c|c|c|c|c|c|c|c|c|r}
\hline\hline
Name   & RA(2000)  &  $A_V$ & $B$&$(B-V)_0$& $SB_{B_0}$& $B_{25}$& $SB_{B_{25}}$ & $Re_B$& $h_B$& $B_e$ & $SB_{B_e}$& $SB_{B_c}$& $R_{min}$       \\
       & Dec.(2000)& $b/a$  & $V$&$(R-I)_0$& $SB_{V_0}$& $V_{25}$& $SB_{V_{25}}$ & $Re_V$& $h_V$& $V_e$ & $SB_{V_e}$& $SB_{V_c}$& $R_{max}$       \\ 
       &           &$V_{LG}$& $R$&         & $SB_{R_0}$& $R_{25}$& $SB_{R_{25}}$ & $Re_R$& $h_R$& $R_e$ & $SB_{R_e}$& $SB_{R_c}$&           \\
       &           &        & $I$&         & $SB_{I_0}$& $I_{25}$& $SB_{I_{25}}$ & $Re_I$& $h_I$& $I_e$ & $SB_{I_e}$& $SB_{I_c}$&           \\ 
  {\tiny (1)}  &    {\tiny (2)}     &   {\tiny (3)}&  {\tiny (4)}  &  {\tiny (5)}   & {\tiny (6)}  &  {\tiny (7)}      & {\tiny (8)}       &  {\tiny (9)}  &  {\tiny (10)} &  {\tiny (11)}   &   {\tiny (12)}       &  {\tiny (13)}  & {\tiny (14)} \\ \hline
UGC05575   & 10 19 46.9&  0.10&  16.04&   0.65& 22.62&   16.43&  23.74&  11.60&  6.91&16.91& 23.32& 22.29&   1.16     \\ 
           & 22 35 42.0&  0.44&  15.38&   0.47& 21.87&   15.60&  23.38&  11.61&  6.92&16.23& 22.64& 21.62&  33.66     \\                                                                                         
           &           & 1356 &  14.97&       & 21.37&   15.08&  23.14&  12.14&  7.24&15.74& 22.21& 21.25&           \\
\hline\hline                                                                                                                          
\end{tabular}                                                                                                                                              
\end{center}                                                                                                                
\end{table}   
 \setcounter{table}{1}
\begin{table}
\begin{center}
\vspace{-1.0cm}
\scriptsize
\caption{Photometric Properties and Velocities of the dIrrs in the Vicinity of the Virgo
Cluster (continued).}
\begin{tabular}{l|c|c|c|c|c|c|c|c|c|c|c|c|r}           
\hline\hline 
Name     & RA(2000)  &  $A_V$  & $B$&$(B-V)_0$& $SB_{B_0}$& $B_{25}$&$SB_{B_{25}}$ & $Re_B$& $h_B$  & $B_e$ & $SB_{B_e}$& $SB_{B_c}$& $R_{min}$       \\
         & Dec.(2000)& $b/a$   & $V$&$(R-I)_0$& $SB_{V_0}$& $V_{25}$&$SB_{V_{25}}$ & $Re_V$& $h_V$  & $V_e$ & $SB_{V_e}$& $SB_{V_c}$& $R_{max}$       \\ 
         &           & $V_{LG}$& $R$&         & $SB_{R_0}$& $R_{25}$&$SB_{R_{25}}$ & $Re_R$& $h_R$  & $R_e$ & $SB_{R_e}$& $SB_{R_c}$&           \\
         &           &         & $I$&         & $SB_{I_0}$& $I_{25}$&$SB_{I_{25}}$ & $Re_I$& $h_I$  & $I_e$ & $SB_{I_e}$& $SB_{I_c}$&           \\  
     {\tiny (1)}  &    {\tiny (2)}     &   {\tiny (3)}&  {\tiny (4)}  &  {\tiny (5)}   & {\tiny (6)}  &  {\tiny (7)}      & {\tiny (8)}       &  {\tiny (9)}  &  {\tiny (10)} &  {\tiny (11)}   &   {\tiny (12)}       &  {\tiny (13)}  & {\tiny (14)} \\ \hline
 PGC2832077& 11 00 40.1&  0.08&  17.42 &  0.42& 24.06&   18.85&  24.75&  11.59& 6.90& 18.32&   25.01& 23.68&    0.92     \\ 
           & 29 42 30.0&  0.52&  17.00 &  0.39& 23.96&   17.89&  24.47&  12.55& 7.48& 17.76&   24.58& 23.33&   22.97     \\                                                                                        
           &           & 616  &  16.76 &      & 23.70&   17.24&  24.30&  13.36& 7.96& 17.37&   24.24& 23.10&            \\
\hline\hline                                                                                                                          
\end{tabular}                                                                                                                                              
\end{center}                                                                                                                
\end{table}   
 \setcounter{table}{1}
\begin{table}
\begin{center}
\vspace{-1.0cm}
\scriptsize
\caption{Photometric Properties and Velocities of the dIrrs in the Vicinity of the Virgo
Cluster (continued).}
\begin{tabular}{l|c|c|c|c|c|c|c|c|c|c|c|c|r} 
\hline\hline 
Name     & RA(2000)  &  $A_V$ &   $B$ &$(B-V)_0$  & $SB_{B_0}$& $B_{25}$&$SB_{B_{25}}$ & $Re_B$& $h_B$  & $B_e$ & $SB_{B_e}$& $SB_{B_c}$& $R_{min}$       \\
         & Dec.(2000)& $b/a$  &   $V$ &$(R-I)_0$  & $SB_{V_0}$& $V_{25}$&$SB_{V_{25}}$ & $Re_V$& $h_V$  & $V_e$ & $SB_{V_e}$& $SB_{V_c}$& $R_{max}$       \\ 
         &           &$V_{LG}$& $R$   &           & $SB_{R_0}$& $R_{25}$&$SB_{R_{25}}$ & $Re_R$& $h_R$  & $R_e$ & $SB_{R_e}$& $SB_{R_c}$&           \\
         &           &        &   $I$ &           & $SB_{I_0}$& $I_{25}$&$SB_{I_{25}}$ & $Re_I$& $h_I$  & $I_e$ & $SB_{I_e}$& $SB_{I_c}$&           \\  
     {\tiny (1)}  &    {\tiny (2)}     &   {\tiny (3)}&  {\tiny (4)}  &  {\tiny (5)}   & {\tiny (6)}  &  {\tiny (7)}      & {\tiny (8)}       &  {\tiny (9)}  &  {\tiny (10)} &  {\tiny (11)}   &   {\tiny (12)}       &  {\tiny (13)}  & {\tiny (14)} \\ \hline
 PGC035087 & 11 25 01.8&  0.09&  17.12&   0.72& 23.48&   17.97&  24.21&  13.58&  8.09& 17.70&   24.42&  23.49&    1.15      \\ 
           & 17 05 09.0&  0.62&  16.40&   0.54& 23.14&   16.90&  24.01&  14.63&  8.72& 17.05&   23.90&  22.98&   28.12      \\                                                                                       
           &           & 1099 &  15.90&       & 22.82&   16.24&  23.84&  15.96&  9.51& 16.51&   23.59&  22.67&             \\
\hline\hline                                                                                                                          
\end{tabular}                                                                                                                                              
\end{center}                                                                                                                
\end{table}   
 \setcounter{table}{1}
\begin{table}
\begin{center}
\vspace{-1.0cm}
\scriptsize
\caption{Photometric Properties and Velocities of the dIrrs in the Vicinity of the Virgo
Cluster (continued).}
\begin{tabular}{l|c|c|c|c|c|c|c|c|c|c|c|c|r} 
\hline\hline 
Name     & RA(2000)  &  $A_V$  & $B$ &$(B-V)_0$& $SB_{B_0}$& $B_{25}$&$SB_{B_{25}}$ & $Re_B$& $h_B$  & $B_e$ & $SB_{B_e}$& $SB_{B_c}$& $R_{min}$       \\
         & Dec.(2000)& $b/a$   & $V$ &$(R-I)_0$& $SB_{V_0}$& $V_{25}$&$SB_{V_{25}}$ & $Re_V$& $h_V$  & $V_e$ & $SB_{V_e}$& $SB_{V_c}$& $R_{max}$       \\ 
         &           & $V_{LG}$& $R$ &         & $SB_{R_0}$& $R_{25}$&$SB_{R_{25}}$ & $Re_R$& $h_R$  & $R_e$ & $SB_{R_e}$& $SB_{R_c}$&           \\
         &           &     &   $I$   &         & $SB_{I_0}$& $I_{25}$&$SB_{I_{25}}$ & $Re_I$& $h_I$  & $I_e$ & $SB_{I_e}$& $SB_{I_c}$&           \\
{\tiny (1)}& {\tiny (2)}  &   {\tiny (3)}&  {\tiny (4)}  &  {\tiny (5)}   & {\tiny (6)}  &  {\tiny (7)}      & {\tiny (8)}       &  {\tiny (9)}  &  {\tiny (10)} &  {\tiny (11)}   &   {\tiny (12)}       &  {\tiny (13)}  & {\tiny (14)} \\ \hline
 PGC3267517& 11 55 58.6&  0.09&  17.69&   0.38& 23.49& 18.43&  24.26&  8.58&  5.11& 18.49&   24.23&  23.34&   0.98       \\ 
           &-03 33 45.0&  0.48&  17.30&   0.13& 23.30& 17.93&  23.94&  8.65&  5.15& 18.10&   23.83&  22.94&  30.89       \\                                                                                      
           &           & 2290 &  17.07&       & 23.04& 17.60&  23.70&  8.67&  5.17& 17.84&   23.57&  22.68&             \\
\hline\hline                                                                                                                          
\end{tabular}                                                                                                                                              
\end{center}                                                                                                                
\end{table}   
 \setcounter{table}{1}
\begin{table}
\begin{center}
\vspace{-1.0cm}
\scriptsize
\caption{Photometric Properties and Velocities of the dIrrs in the Vicinity of the Virgo
Cluster (continued).}
\begin{tabular}{l|c|c|c|c|c|c|c|c|c|c|c|c|r}
\hline\hline 
Name     & RA(2000)  &  $A_V$ & $B$ &$(B-V)_0$ & $SB_{B_0}$& $B_{25}$&$SB_{B_{25}}$ & $Re_B$& $h_B$  & $B_e$ & $SB_{B_e}$& $SB_{B_c}$& $R_{min}$       \\
         & Dec.(2000)& $b/a$  & $V$ &$(R-I)_0$ & $SB_{V_0}$& $V_{25}$&$SB_{V_{25}}$ & $Re_V$& $h_V$  & $V_e$ & $SB_{V_e}$& $SB_{V_c}$& $R_{max}$       \\ 
         &           &$V_{LG}$& $R$ &          & $SB_{R_0}$& $R_{25}$&$SB_{R_{25}}$ & $Re_R$& $h_R$  & $R_e$ & $SB_{R_e}$& $SB_{R_c}$&           \\
         &           &        & $I$ &          & $SB_{I_0}$& $I_{25}$&$SB_{I_{25}}$ & $Re_I$& $h_I$  & $I_e$ & $SB_{I_e}$& $SB_{I_c}$&           \\  
  {\tiny (1)}  &    {\tiny (2)}     &   {\tiny (3)}&  {\tiny (4)}  &  {\tiny (5)}   & {\tiny (6)}  &  {\tiny (7)}      & {\tiny (8)}       &  {\tiny (9)}  &  {\tiny (10)} &  {\tiny (11)}   &   {\tiny (12)}       &  {\tiny (13)}  & {\tiny (14)} \\ \hline
 PGC038872 & 12 12 04.5&  0.08 & 17.37 &  0.57& 23.90 &  18.61 & 24.66 &12.71 &     7.57 & 18.08 &  25.06 &  23.88 &   0.99      \\ 
           & 12 44 08.0&  0.41 & 16.80 &  0.56& 23.46 &  17.55 & 24.39 &13.55 &     8.07 & 17.42 &  24.52 &  23.34 &  32.08      \\                                                                                       
           &           & 1496  & 16.44 &      & 23.05 &  16.84 & 23.87 &13.86 &     8.26 & 16.97 &  23.79 &  22.87 &            \\
\hline\hline                                                                                                                          
\end{tabular}                                                                                                                                              
\end{center}                                                                                                                
\end{table}   
  \setcounter{table}{1}
\begin{table}
\begin{center}
\vspace{-1.0cm}
\scriptsize
\caption{Photometric Properties and Velocities of the dIrrs in the Vicinity of the Virgo
Cluster (continued).}
\begin{tabular}{l|c|c|c|c|c|c|c|c|c|c|c|c|r}
\hline\hline 
Name     & RA(2000)  &  $A_V$  & $B$&$(B-V)_0$& $SB_{B_0}$& $B_{25}$&$SB_{B_{25}}$ & $Re_B$& $h_B$  & $B_e$ & $SB_{B_e}$& $SB_{B_c}$& $R_{min}$       \\
         & Dec.(2000)& $b/a$   & $V$&$(R-I)_0$& $SB_{V_0}$& $V_{25}$&$SB_{V_{25}}$ & $Re_V$& $h_V$  & $V_e$ & $SB_{V_e}$& $SB_{V_c}$& $R_{max}$       \\ 
         &           & $V_{LG}$& $R$&         & $SB_{R_0}$& $R_{25}$&$SB_{R_{25}}$ & $Re_R$& $h_R$  & $R_e$ & $SB_{R_e}$& $SB_{R_c}$&           \\
         &           &         & $I$&         & $SB_{I_0}$& $I_{25}$&$SB_{I_{25}}$ & $Re_I$& $h_I$  & $I_e$ & $SB_{I_e}$& $SB_{I_c}$&           \\  
{\tiny {\tiny (1)}}  &    {\tiny (2)}     &   {\tiny (3)}&  {\tiny (4)}  &  {\tiny (5)}   & {\tiny (6)}  &  {\tiny (7)}      & {\tiny (8)}       &  {\tiny (9)}  &  {\tiny (10)} &  {\tiny (11)}   &   {\tiny (12)}       &  {\tiny (13)}  & {\tiny (14)} \\ \hline
 CS1008    & 12 16 51.4 & 0.07&  17.02 &  0.77& 22.69&   17.79&  24.01&  12.11&  7.22& 17.68&   24.12& 23.23&    1.24     \\ 
           & 29 03 17.0 & 0.78&  16.25 &  0.52& 22.12&   16.68&  23.72&  11.78&  7.02& 17.00&   23.38& 22.48&   37.77     \\                                                                                        
           &            &1100 &  15.75 &      & 21.93&   16.02&  23.49&  11.49&  6.85& 16.52&   22.90& 21.95&            \\
\hline\hline                                                                                                                          
\end{tabular}                                                                                                                                              
\end{center}                                                                                                                
\end{table}   

\end{document}